\documentclass[12pt,preprint]{aastex}

\shorttitle{K-shell X-ray Lines from Oxygen Ions}
\shortauthors{Gu et al.}
\received{}
\revised{}
\accepted{}
\ccc{AAS}

\begin{document}

\title{Laboratory Measurement and Theoretical Modeling of K-shell X-ray Lines
   from Inner-shell Excited and Ionized Ions of Oxygen}
\author{
Ming Feng Gu\altaffilmark{1},
Mike Schmidt\altaffilmark{2,3},
Peter Beiersdorfer\altaffilmark{2},
Hui Chen\altaffilmark{2},
Daniel B. Thorn\altaffilmark{2},
Elmar Tr\"{a}bert\altaffilmark{2,4},
Ehud Behar\altaffilmark{5}, and
Steven M. Kahn\altaffilmark{1}
}
\altaffiltext{1}{Kavli Institute for Particle Astrophysics and Cosmology,
and \\ Department of Physics, Stanford University, CA 94305, USA}
\altaffiltext{2}{High Temperature and Astrophysics Division,
Physics and Advanced Technologies, \\ Lawrence
Livermore National Laboratory, Livermore, CA 94550-9234, USA}
\altaffiltext{3}{Technische Universit\"{a}t Dresden, Germany}
\altaffiltext{4}{Ruhr-Universit\"{a}t Bochum, Germany}
\altaffiltext{5}{Department of Physics, Technion, Haifa 32000, Israel}

\begin{abstract}
We present high resolution laboratory spectra of K-shell X-ray lines from
inner-shell excited and ionized ions of oxygen, obtained with a reflection
grating spectrometer on the electron beam ion trap (EBIT-I) at the Lawrence
Livermore National Laboratory. Only with a multi-ion model including all major
atomic collisional and radiative processes, are we able to identify the
observed K-shell transitions of oxygen ions from \ion{O}{3} to \ion{O}{6}. The
wavelengths and associated 
errors for some of the strongest transitions are given, taking into
account both the experimental and modeling uncertainties. The present data
should be useful in identifying the absorption features present in
astrophysical sources, such as active galactic nuclei and X-ray
binaries. They are also useful in providing benchmarks for the testing of
theoretical atomic structure calculations.
\end{abstract}


\section{INTRODUCTION}

The powerful diffraction grating instruments on board {\it Chandra}
\citep{Canizares00,Brinkman00} and {\it XMM-Newton}
\citep{den Herder01} have enabled the observation of K-shell X-ray absorption
lines of the ionized plasma surrounding Active Galactic Nuclei (AGNs).
Absorption features attributed to various ionization stages of oxgygen
ranging from neutral to as high as \ion{O}{8} have been seen,
for example, in NGC 5548, NGC 3783, NGC 4593, and NGC 7469
\citep{Behar03, Steenbrugge03a, Steenbrugge03b, Blustin03}.
These new K-shell X-ray absorption measurements complement
those in the ultraviolet band (e.g., \citet{Arav03, Crenshaw03}).

One important physical property that can be derived from such
measurements is the outflow velocity, which is based on the
shift of a given line from its wavelength at rest. Clearly,
the accuracy of this procedure depends strongly on the accuracy
with which the rest frame wavelength is known  \citep{Behar02}.
Calculations offer only limited guidance, as calculations have a
rather high uncertainty. A comparison showed that K-shell X-ray line
positions of \ion{O}{6} and \ion{O}{5} calculated by several
authors differ by as much as 60~m{\AA} and 80~m{\AA}, respectively
\citep{Schmidt04}. These differences correspond to uncertainties
between 800 and 1100~km~s$^{-1}$, respectively. The spread in the
calculated rest frame wavelengths is thus significantly larger
than the wavelength shift expected for typical outflow velocities,
making most theoretical values nearly useless as rest frame reference
standards for a given line. Because of the difficulty of accurately predicting
electron-electron correlations, the uncertainty is expected to
increase further for calculations of the K-shell X-ray lines of
the lower charge states of oxygen. Only laboratory measurements
can establish the rest frame wavelengths of a given line and provide
an associated uncertainty estimate.

Recently, \citet{Schmidt04} reported a laboratory measurement of the rest
frame wavelengths of the K-shell X-ray resonance lines in \ion{O}{6} and
\ion{O}{5}. The experimental precision was sufficient to determine flow
velocities to within 20 -- 40~km~s$^{-1}$, exceeding the typical measurement
accuracy of any current astrophysical X-ray instrument. These measurements
were conducted by emission spectroscopy in the interaction of oxygen ions with
electrons of sufficient energy to excite the respective K-shell
transition. Because the energy needed to excite a K-shell transition
is significantly larger than the energy required to ionize all but
heliumlike and hydrogenlike oxygen, the measurements had to be
performed in a non-equilibrium, ionizing plasma.

Here we report on measurements of K-shell X-ray lines from charge
states of oxygen as low as \ion{O}{3}. These measurements again
utilize emission spectroscopy of oxygen ions excited by high-energy
electrons in a non-equilibrium, ionizing plasma. The lines observed
under such conditions may be similar to those observed in a photoionized
spectrum, where ionization by an electron is replaced by ionization
by X rays.

The interpretation of non-equilibrium, ionizing plasmas is difficult.
This has recently been shown in various analyses of K-shell iron
spectra \citep{Decaux97, Decaux03, Jacobs97}.
Difficulties arise (1) because the wavelengths of the lines are not
well known so that the line assignment may be uncertain when
clusters of lines are involved and (2) because the excitation
processes may be complex, involving an interplay of various processes
that are often not fully modeled (if at all),
e.g., inner-shell ionization, inner-shell excitation, autoionization,
and radiative cascades.

In the simplest case, it is necessary to calculate all excitation,
radiative, and autoionization rates affecting a given charge state. Because
levels in two neighboring charge states are involved in the formation of
spectra due to the presence of autoionization transitions, we denote this
model as a ``two-ion'' model. This model is insufficient to predict the
emission in an
ionizing plasma where, for example, inner-shell ionization of the
neighboring, lower charge state can produce an excited state that
in turn contributes to the observed spectrum when undergoing radiative
decay. A complex ``three-ion" model is typically used to describe the
emission from plasmas in which significant ionization and recombination
processes take place. Such a model was used, for example, by \citet{Doron02}
to examine the effect of ionization of \ion{Fe}{16} and the recombination
from \ion{Fe}{18} on the \ion{Fe}{17} X-ray emission. We show that
the three-ion model is also insufficient to describe the K-shell
oxygen emission observed in our ionizing plasma. In order to describe
our observations, we present an ``all-ion" model. In this model,
the emission of a given ion is coupled to more than just its nearest
neighbors. Only this model is found to be able to satisfactorily
reproduce the observed oxygen K-shell emission.

The combination of the laboratory data and the all-ion model
allows us to identify lines from \ion{O}{3}, \ion{O}{4},
\ion{O}{5}, and \ion{O}{6} that, to the best of our knowledge,
have not yet been measured before. Although many features are
likely to be blends of several lines, the features are sufficiently
narrow so that we can readily assign wavelengths with small
uncertainties. In addition to aiding the identification of lines
in absorption spectra, the measured line positions are useful for
establishing the quality of wavelength computations and as empirical
input for optimizing calculational schemes, such as those carried out recently
by \citet{Garcia04}.

\section{MEASUREMENT}

The measurement was performed at the University of California Lawrence
Livermore National Laboratory using the EBIT-I electron beam ion trap.
The device has been used for laboratory astrophysics measurements for
over a decade \citep{Beiersdorfer03}.

The X-ray emission of EBIT-I was detected by a high-resolution grazing
incidence spectrometer \citep{Beiersdorfer04} with an angle
of incidence of 88.5$^{\circ}$.
The spectrometer features a variable line-spaced reflective grating with
an average line spacing of 2400~grooves/mm and a radius of curvature
of 44.3~m. The spectral image of the ion cloud in the ion trap
is nearly flat and permits the use of a
two-dimensional charge-coupled device (CCD) as a multichannel detector.
The detector is cooled by liquid nitrogen and employs a
27~mm~$\times$~26~mm, thinned,
back-illuminated CCD chip with 1340~$\times$~1300 pixels of
20~$\mu$m $\times$ 20~$\mu$m nominal size; the quantum efficiency is
about 45\% at 22~{\AA}.
In the present setting the detector spans a wavelength interval about
7~{\AA} wide,
which is ample to cover the present 21-24~{\AA} range of interest. The
resolving power of this setup, drawn from the measured spectra, is about 1100.

Oxygen was introduced to the electron beam ion trap by injection
of carbon dioxide. This option
provided {\it in situ} calibration lines in form of helium-like oxygen
(\ion{O}{7}) as well as hydrogen-like carbon (\ion{C}{6}). We observed the
resonance, intercombination, and forbidden lines of \ion{O}{7}, denoted $w$,
$y$, and $z$, respectively, following the labeling convention introduced by
\citet{Gabriel72}.  The wavelengths of these lines are well known from
the calculations of \citet{Drake88} and some also from measurements by
\citet{Engstroem95}.
We also observed the {\sc Lyman} series in the hydrogen-like spectrum of
\ion{C}{6}, {$\rm Ly\beta$} through {$\rm Ly\varepsilon$}, for which
\citet{GarciaMack64} provide calculated wavelengths. All these calibration
line wavelengths are presumed to be accurate to better than 1~m{\AA}.

\begin{figure}
\epsscale{0.7}
\plotone{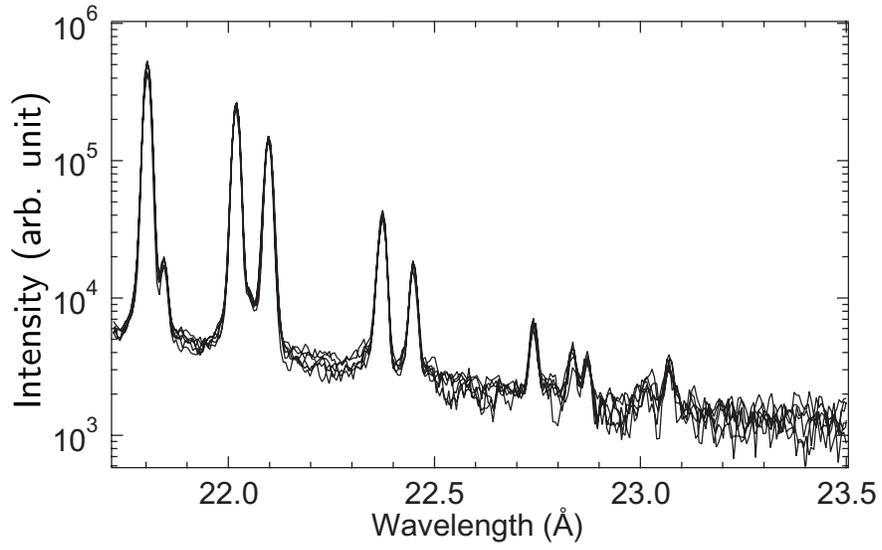}

\caption{\label{fig:mspec} Ten superimposed spectra of oxygen K-shell
  lines observed in the wavelength range of 21.7 -- 23.5~{\AA}. Each spectrum
  is the result of 120 min observation.}
\end{figure}

Figure~\ref{fig:mspec} shows features from 21.7--23.5~{\AA}. Ten superimposed
spectra, each with 120 min exposure time, show the reproducibility of weak
emission features. The black trace in Figure~\ref{fig:spec} 
represents the sum of the ten spectra in the wavelength range of 21.4 --
23.2~{\AA}, from 20~h of total observation time.
Each of the ten spectra was individually filtered against cosmic ray events, 
calibrated, and analyzed. The reproducibility of these ten measurements
was used to assess statistical and systematic errors.

The measured spectrum in Figure~\ref{fig:spec} is dominated by the He-like
lines $w$, $y$, and $z$. We also readily identify the Li-like line pair $q$
and $r$. These constitute the four strongest lines seen in the spectrum. In
addition, the spectrum contains nine features, labeled $A$ -- $I$. These
satellite lines belong to charge states between \ion{O}{3} and \ion{O}{6}. They
are situated at the long-wavelength side of the spectrum, and
their intensities are considerably weaker than those of the \ion{O}{7}
lines. This is partly due to the fact that the upper levels of these
transitions have high autoionization rates, and therefore small radiative
branching ratios. Another reason lies in the ionization balance; at the
electron energies that are needed to excite K-shell lines, the charge state
distribution strongly favors \ion{O}{7},
leaving only a small fraction of \ion{O}{6} and a minute abundance of
\ion{O}{5} and lower charge states (\ion{O}{4} and \ion{O}{3}).
In fact, the ionization potential of \ion{O}{4} and \ion{O}{3} is about
114~eV and 77~eV, respectively, while for
technical reasons we used an electron beam energy of 4~keV. We therefore
continually supplied neutral oxygen and then detected the different charge
states lines (\ion{O}{3} through \ion{O}{6}) during the ionizing phase,
before the ionization equilibrium was reached.
This procedure is similar to that used in earlier measurements of the
K-shell emission of low charge state ions of iron \citep{Decaux95,Decaux97}.

The wavelengths of all features, $A$ -- $I$,
are determined relative to the reference lines from \ion{O}{7} and \ion{C}{6},
and are tabulated in Table~\ref{tab:id}. The uncertainty of the reference
lines includes two contributions, one from the errors of their observed
positions and the other from their wavelength errors.
The uncertainty of reference line positions was added in quadrature
to the uncertainty of the line positions of the measured line features. The
uncertainty of the reference line wavelengths, which is assumed to be less
than 1~m{\AA}, was added linearly.
The overall wavelength uncertainty ranges from 2 to 4~m{\AA} for line features
$C$ -- $G$ and $I$, and about 10 -- 15~m{\AA} for $A$, $B$ and $H$.

It is not trivial to unambiguously identify the 
emission features $A$ -- $I$ with
specific transitions in various oxygen ions. After examining several
theoretical models, we find that only a complex multi-ion
model including major collisional and radiative processes is able to
account for all line features. In the following we explain the details of these
theoretical models, and the most likely identifications of the observed emission
lines.

\section{THEORETICAL MODELS}
The modeling of K-shell emission spectra of L-shell ions under the
EBIT-I plasma
conditions requires the inclusion of various atomic processes, such
as electron collisional
excitation, radiative decay, autoionization, and direct electron collisional
ionization. With the present beam energy, recombination processes are
generally insignificant in populating the excited levels. Because no external
radiation field is present in EBIT, we do not include photoionization process
in the model.
The Flexible Atomic Code (FAC) is used in
the present analysis for the basic atomic parameters. FAC is a relativistic
configuration interaction atomic code, and uses the distorted-wave
approximation to treat electron-ion collision processes \citep{Gu03}.

With the given non-equilibrium charge balance for O ions, inner-shell
direct electron collisional ionization is
expected to be a major process in forming the K-shell lines in addition to the
collisional excitation. Moreover,  with an electron density of
$\approx 10^{12}$~cm$^{-3}$ provided by the electron beam current of 120~mA in
EBIT-I, many levels in the $1s^22l^q$ configurations are
significantly populated, and inner-shell excitation as well as ionization from
these excited states also play important roles in forming the observed
spectrum. To illustrate these effects, we have constructed three theoretical
collisional-radiative models with increasing sophistication, which we denote
as two-ion, 
three-ion, and all-ion models. The electron density used in these models,
$1.2\times 10^{12}$~cm$^{-3}$, is determined by matching the observed
\ion{O}{7} line ratio $y/z$, which is density sensitive,
to the theoretical result of the all-ion model. In all three models, we only
consider the atomic states belonging to the $1s^22l^q$, $1s^22l^{q-1}nl$ and $1s2l^qnl$
($2\le n\le 4$) configuration complexes for each ion. In 
the calculation of atomic data, configuration interaction within the same
complex is included.

\begin{figure}
\epsscale{0.7}
\plotone{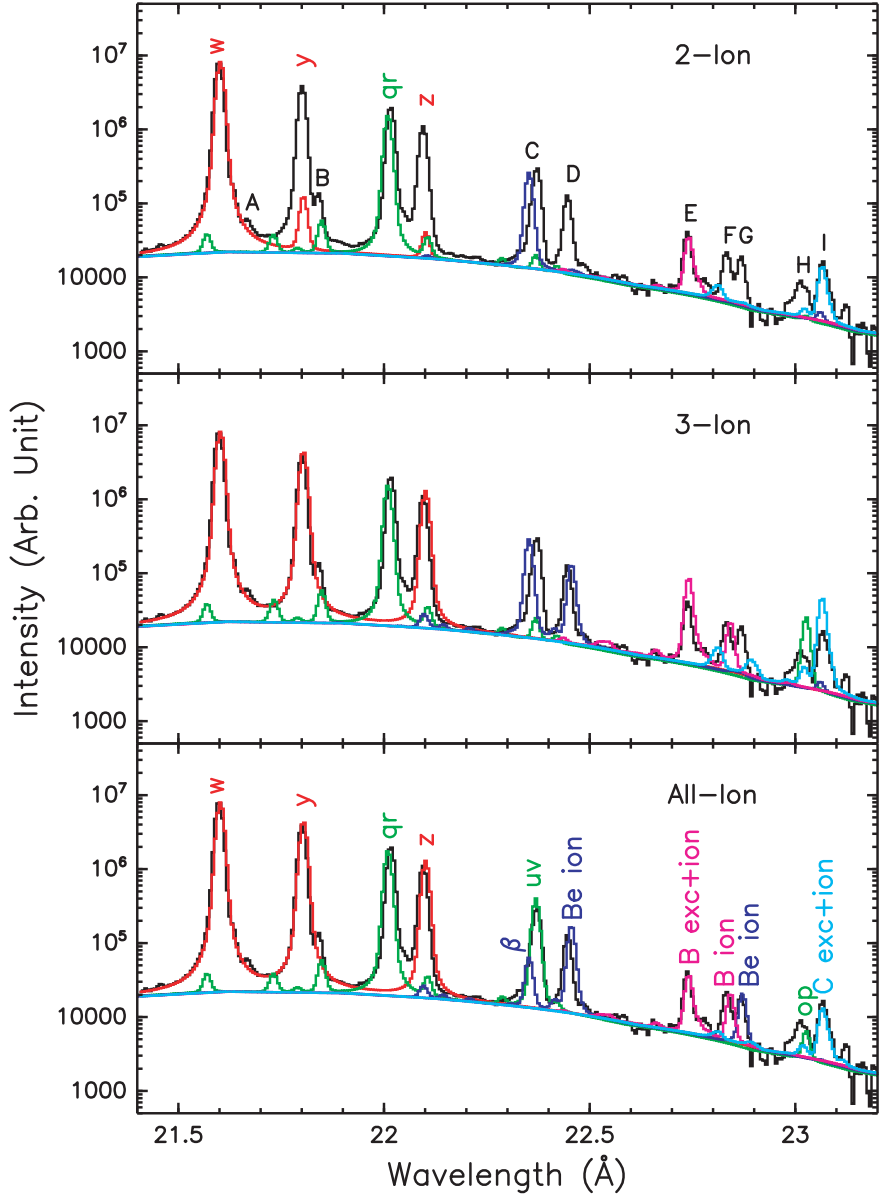}

\caption{\label{fig:spec}Comparison of 2-ion, 3-ion, and all-ion models
with the observed spectrum. Black: experiment, red: \ion{O}{7} lines,
green: \ion{O}{6}
    lines, blue: \ion{O}{5} lines, magenta: \ion{O}{4} lines, and cyan:
\ion{O}{3} lines. The
    theoretical wavelengths in the figure are the configuration interaction
    results obtained with FAC. The \ion{O}{6} lines have been shifted to
    the left by 28~m{\AA}, and the \ion{O}{5} lines have been shifted to
    the left by 10~m{\AA}. Labels $A$ -- $I$ serve to identify the spectral
    features discussed in the text and listed in Table~\ref{tab:id}. }
\end{figure}

In the two-ion model, we ignore the direct electron collisional ionization
processes, and only include collisional excitation followed by 
radiative cascades and autoionization
transitions into the next higher charge state. Therefore, only levels of two
neighboring charge states are involved in determining the spectrum of a
particular ion. The emission lines from \ion{O}{3} -- \ion{O}{7} are
calculated separately,
with the fractional abundance of each charge state adjusted to match the
brightest lines observed. The result of this model is shown in the top panel
of Figure~\ref{fig:spec}. The relative abundances of \ion{O}{3} --
\ion{O}{7} ions are
0.36, 0.42, 1.44, 0.77, 1.00, respectively. With this model, it is clear that
the intensities of \ion{O}{7} lines $y$ and $z$, and the line features
labeled as $D$, $F$, $G$, and $H$ cannot be explained. Feature $C$ is
attributed to the \ion{O}{5} line $\beta $, and features $E$ and $I$ are
attributed to the \ion{O}{4}
and \ion{O}{3} lines, respectively. Features $A$ and $B$ do not seem to
belong to the KLL satellite transitions, and their nature is discussed later
in this section. The failure of this model is due to the lack of ionization
processes that make contributions to certain lines.

In the three-ion model, we add the next lower 
charge state to the modeling of the
spectrum of a particular ion, and therefore we include the inner-shell
electron collisional ionization processes
forming K-shell lines in addition to the collisional excitation and
autoionization. However, the spectral contributions from 
individual ions are
again treated in separate calculations, with the fractional abundance of each
charge state adjusted to match the brightest lines, which are the same as
those used in the two-ion model, except that one more ion, \ion{O}{2} with a
fractional abundance of 0.12 is added to the spectral model of \ion{O}{3}. The
result of this model is shown in the middle panel of
Figure~\ref{fig:spec}. The three-ion model represents a significant
improvement over the two-ion model. The \ion{O}{7} lines $y$ and $z$ and the
line features D and F are now
accounted for satisfactorily. However, feature $H$, which is attributed to
the \ion{O}{6} lines $o$ and $p$, is now overpredicted. Lines $o$ and $p$ are
mainly produced by the
ionization of \ion{O}{5}, whose abundance is fixed by matching the
$\beta$ line, which is mainly produced by collisional excitation, to the line
feature $C$. Therefore the ratio of 
($o+p$)/$\beta $ in this model does not depend
on the relative abundances of the \ion{O}{6} and \ion{O}{5} ions, and is in
disagreement with the observed spectrum. 
Moreover, the observed line feature $G$
cannot be explained satisfactorily in this model. It can be tentatively
attributed to the \ion{O}{3}
lines in its vicinity, which are mainly due to the ionization of \ion{O}{2}
ions. However, the ionization of \ion{O}{2} ions also contributes to
the intensity of line feature $I$, and with the chosen abundance of 0.12 for
\ion{O}{2}, line $I$ is already
overpredicted, while the \ion{O}{3} lines in the vicinity of $G$ are
not nearly strong enough to account for its intensity.

The problems with the three-ion model arise from the fact that only
levels in three adjacent charge states are allowed to influence the spectrum
of the ion. For example,
when calculating the spectrum of \ion{O}{5}, only levels in \ion{O}{6},
\ion{O}{5}, and \ion{O}{4} enter the model. At very low electron
densities, where only the
ground state of each charge state is significantly populated, this is a good
approximation. At the present electron density, many states in the $1s^22l^q$
configurations have large populations relative to the ground state. The
ionization of these excited states has an important role in populating the
excited states of the next higher charge state, which in turn makes significant
contributions to the K-shell lines of even higher charge states. In the
example of \ion{O}{5}, therefore, all ions having 
lower charge than \ion{O}{5} can
indirectly influence the level population of \ion{O}{5} through successive
ionization. With a three-ion model, such indirect effects are lost. In order
to address such far-reaching line formation processes, we have constructed an
all-ion model, which includes \ion{O}{1}
-- \ion{O}{7} charge states in a single collisional radiative model. The relative
abundances of
\ion{O}{1} -- \ion{O}{7} are adjusted to be 0.01, 0.03, 0.10, 0.15,
0.36, 0.77, and 1.00, respectively, in order to match the brightest lines in
the observed spectrum. The small abundances of \ion{O}{1} and \ion{O}{2} are
required by the lack of any
significant lines at wavelengths longer than 23.2 {\AA} (see
Figure~\ref{fig:mspec}). The result of this 
model is shown in the bottom panel of Figure~\ref{fig:spec}. Unlike the
two and three-ion models, this model clearly accounts for all major emission
lines in the observed spectrum. An interesting 
fact is that now line feature $C$
is mainly attributed
to the \ion{O}{6} lines $u$ and $v$ instead of \ion{O}{5} $\beta $, and
feature $G$ is attributed to \ion{O}{5} lines, both of which arise
due to the step-wise ionization processes from the lower charge states.

The observed spectrum also shows two significant features, $A$ and $B$, on the
long wavelength shoulder of the $w$ and $y$ 
lines. The theoretical models indicate
that all $1s^22l-1s2l^2$ transitions of \ion{O}{6}, notably, $s$, $t$, $n$,
and $m$, are too
weak to explain them, and no lines from lower charge states appear in this
region. However, the models do give several \ion{O}{6} lines resulting from the
$1s^23l^\prime-1s2s3l$ ($l^\prime=l\pm 1$) transitions in the vicinity of this
region, and they have intensities comparable to those observed. These lines are
mainly produced
by the inner-shell excitation of the \ion{O}{6} ground state, and are
enabled only
through configuration interaction effects, since the radiative decay involves
two-electron transitions. However, due to the difficulties in calculating
the radiative and autoionization rates associated with these levels,
the confidence in these identifications is somewhat limited.

In Table~\ref{tab:id}, we list possible line identifications of all
significant features from \ion{O}{3} -- \ion{O}{6} in the observed
spectrum. All of
these features consist of multiple closely spaced lines, and the theoretical
relative intensities of individual lines are given according to the all-ion
model. We also compare the measured wavelengths with various theoretical
calculations and astrophysical observations, wherever available. In the
present calculation, wavelengths are obtained with the configuration
interaction (CI) method for all ions, and with a
combined CI and second-order many-body perturbation theory method (MBPT) for
the KLL transitions of \ion{O}{5} and \ion{O}{6} \citep{Gu04}.
A second-order MBPT calculation is found to 
have no improvements over the CI method
for charge states lower than \ion{O}{5}, and therefore the corresponding
results are not shown. For \ion{O}{5} and \ion{O}{6}, the MBPT correction
does seem to improve the theoretical wavelengths significantly.

\section{DISCUSSION}

Based on the identifications presented in the previous section, we assign
measured wavelengths and associated uncertainties to one or two strongest
lines in each blended feature, as listed in Table~\ref{tab:lines}. The
uncertainties not only reflect the measurement errors, but also take into
account the fact that each observed feature contains contributions from weaker
lines.

One result of this modeling effort is the recognition that
what we had taken to represent line $\beta $ of \ion{O}{5} at 22.374(3)~{\AA}
\citep{Schmidt04} is now seen
as a contribution to a line blend that is dominated by the \ion{O}{6}
satellite lines $u$ and $v$. The line blend has the small wavelength uncertainty
determined previously \citep{Schmidt04}. The individual, unresolved line
constituents now have to be
assigned larger errors. As we do not see any distortion of the joint line
profile from the instrumental profile, line $\beta $ may be assumed to lie
within the half width of the line, giving it a line position uncertainty of
about 30~m{\AA}, or ten times more than that of the full line. By comparing
the model calculations with the observed spectra, we notice that our new MBPT
wavelenghts for other \ion{O}{5} and \ion{O}{6} lines agree very well with the
measured values. It is therefore reasonable to expect that the MBPT wavelength
for line $\beta$ is reliable to within 10~m{\AA}, and we assign it a wavelength
of 22.370(10)~{\AA}. This uncertainty of 10~m{\AA} is about three times larger
than previously assumed, but still significantly better than that of
{\it ab initio} calculations. The uncertainty corresponds to outflow
velocity uncertainties of about 130~km~s$^{-1}$ and is comparable or better than
the measurement uncertainty of any current astrophysical measurement.

Line feature $D$ is mainly comprised of six \ion{O}{5} transitions with
lower and upper configurations of  $1s^22s2p$ and $1s2s2p^2$, respectively. Of
these six lines, the $J=2\to J=2$ transition is the strongest with
roughly 40\% of the total intensity. All six transitions have theoretical
wavelengths within 4~m{\AA} of each other. Based 
on the representative wavelength measurement of the complex, we assign an 
uncertainty of 8~m{\AA} to the lines in
Table~\ref{tab:lines}.

Feature $E$ has two dominant components with 80\% of the total
intensity. The two transitions have similar 
theoretical wavelengths. We therefore
expect the measured wavelength to be an accurate representation of these two
transitions, and we assign a small uncertainty of 5~m{\AA} to them.

Feature $F$ is almost entirely due to two \ion{O}{4} transitions
between $1s^22s2p^2$($\frac{5}{2}$,$\frac{3}{2}$) and
$1s2s2p^3$($\frac{3}{2}$) states. The two transitions have very close
theoretical wavelengths, thereofore, we assign the measured wavelength to this
blended feature with a small uncertainty.

Feature $G$ is mainly due to the \ion{O}{5} transition $1s^22s2p$(2) --
$1s2s2p^2$(2) with 75\% of the total intensity. The remaining blending have
the calculated wavelengths very close to the main component. Accordingly, a
small uncertainty is assigned to the main component.

Feature $H$ is mainly due to the \ion{O}{6} line pair $o$ and
$p$. However, this feature is quite weak and broad. The calculated intensity
of $o$ and $p$ does not agree with the observed intensity as well as
other features do. It
is possible that many weak transitions from other charge states contribute to
this feature. We therefore assign a relatively large uncertainty of 20~m{\AA}
to this line pair.

The major component of feature $I$ is the blend of the \ion{O}{3} lines
$1s^22s^22p^2$(1,2) -- $1s2s^22p^3$(1) with 70\% of the total intensity, and
having nearly identical theoretical wavelengths. We therefore assign the
measured wavelength of this feature as belonging 
to these two transitions, with a
small uncertainty of 6~m{\AA}.

The wavelengths determined for these transitions in the present work are all
significantly better than what is achievable with {\it ab initio}
calculations, and are better than or at least comparable to the accuracy of the
\textit{Chandra} and \textit{XMM-Newton} grating instruments. Our data provide
reliable reference lines for outflow velocity measurements in the absorption
spectroscopy of AGNs and X-ray binaries.

\acknowledgments
The work at Lawrence Livermore National Laboratory was performed under
the auspices of the Department of Energy under Contract No. W-7405-Eng-48
and was supported by the National Aeronautics and Space
Administration under work order W19,878 issued by the Space
Astrophysics Research and Analysis Program. M.F.G and S.M.K acknowledge the
support by the NASA grant NAG5-5419.
E.T. acknowledges travel support from the German Research Association (DFG).
E.B. was supported by grant No. 2002111 from the United States Israel
Binational Foundation.

\clearpage

\newpage
\begin{deluxetable}{ *{14}{c} }
\rotate
\tabletypesize{
\scriptsize
}
\tablecaption{\label{tab:id}Line identifications and wavelength comparisons
    for K-shell lines of \ion{O}{3} -- \ion{O}{6}.}
\tablehead{
\colhead{Index} &
\colhead{$\lambda_{\mbox{exp}}$\tablenotemark{a}} &
\colhead{Ion} &
\colhead{Lower($J$)} &
\colhead{Upper($J$)} &
\colhead{$f_{ij}$\tablenotemark{b}} &
\colhead{$\lambda_{\mbox{CI}}$\tablenotemark{c}} &
\colhead{$\lambda_{\mbox{MBPT}}$\tablenotemark{d}} &
\colhead{$\lambda_{\mbox{G}}$\tablenotemark{e}} &
\colhead{$\lambda_{\mbox{VS}}$\tablenotemark{f}} &
\colhead{$\lambda_{\mbox{C}}$\tablenotemark{g}} &
\colhead{$\lambda_{\mbox{BK}}$\tablenotemark{h}} &
\colhead{$\lambda_{\mbox{K}}$\tablenotemark{i}} &
\colhead{Intensity}
}
\startdata
$A$ & 21.672(15) & \ion{O}{6} & $1s^23s$($\frac{1}{2}$) &
$1s2p3d$($\frac{3}{2}$) & 1.4[-2] & 21.600 & & & & & & & 0.19 \\
   &            & \ion{O}{6} & $1s^23d$($\frac{5}{2}$) &
$1s2p3d$($\frac{3}{2}$) & 4.5[-3] & 21.763 & & & & & & & 0.18 \\
$B$ & 21.845(10) & \ion{O}{6} & $1s^23s$($\frac{1}{2}$) &
$1s2s3p$($\frac{3}{2}$) & 1.6[-2] & 21.846 & & & & & & & 0.58 \\
   &            & \ion{O}{6} & $1s^23s$($\frac{1}{2}$) &
$1s2s3p$($\frac{1}{2}$) & 7.5[-3] & 21.846 & & & & & & & 0.27 \\
$C$ & 22.374(3) & \ion{O}{6} & 
$1s^22s$($\frac{1}{2}$) & $1s2s2p$($\frac{1}{2}$) 
 & 2.4[-6] & 22.397 & 22.377 & 22.37& & 22.42
& & 22.377(12) & 5.49 \\
                &        & \ion{O}{6} & $1s^22s$($\frac{1}{2}$)&
$1s2s2p$($\frac{3}{2}$) & 1.2[-5] &
                            22.397 & 22.376 & &22.373 & & & & 4.18 \\
            &        & \ion{O}{5}  & $1s^22s^2$(0) & $1s2s^22p$(1) &
                       5.5[-1] & 22.363 & 22.374 & & 
22.41 & 22.33 & 22.35 & & 1.25 \\
$D$ & 22.449(3) & \ion{O}{5} & $1s^22s2p$(2) & $1s2s2p^2$(2) & 2.0[-1] & 22.465 &
22.451 & & & & & 22.501(19) & 1.65 \\
    &        & \ion{O}{5} & $1s^22s2p$(1) & $1s2s2p^2$(2) & 1.2[-1] & 22.464 &
22.449 & & & & & & 0.60 \\
      &        & \ion{O}{5} & $1s^22s2p$(2) & $1s2s2p^2$(1) & 6.6[-2] & 22.467 &
22.453 & & & & & & 0.54 \\
      &        & \ion{O}{5} & $1s^22s2p$(0) & $1s2s2p^2$(1) & 2.8[-1] & 22.465 &
22.451 & & & & & & 0.46 \\
      &        & \ion{O}{5} & $1s^22s2p$(1) & $1s2s2p^2$(0) & 9.0[-2] & 22.466 &
22.451 & & & & & & 0.45 \\
      &        & \ion{O}{5} & $1s^22s2p$(1) & $1s2s2p^2$(1) & 6.3[-2] & 22.467 &
22.452 & & & & & & 0.32 \\
$E$  & 22.741(4) & \ion{O}{4} & $1s^22s^22p$($\frac{3}{2}$) &
$1s2s^22p^2$($\frac{3}{2}$) & 2.0[-1] & 22.741 & & & 22.75 & 22.73 & 22.73
& 22.740(20) & 0.42 \\
      & & \ion{O}{4} &  $1s^22s^22p$($\frac{1}{2}$) &
$1s2s^22p^2$($\frac{1}{2}$) & 1.6[-1] &
                  22.741 & & & & & & & 0.19 \\
      & & \ion{O}{4} &  $1s^22s^22p$($\frac{3}{2}$) &
$1s2s^22p^2$($\frac{1}{2}$) & 4.0[-2] &
                  22.743 & & & & & & & 0.09 \\
      & & \ion{O}{4} &  $1s^22s^22p$($\frac{1}{2}$) &
$1s2s^22p^2$($\frac{3}{2}$) & 6.0[-2] &
                  22.739 & & & & & & & 0.06 \\
$F$  & 22.836(4) & \ion{O}{4} & $1s^22s2p^2$($\frac{5}{2}$) &
$1s2s2p^3$($\frac{3}{2}$) & 1.0[-1] & 22.843 & & & & & & & 0.20 \\
      & & \ion{O}{4}          & $1s^22s2p^2$($\frac{3}{2}$) &
$1s2s2p^3$($\frac{3}{2}$) & 1.1[-1] & 22.842 & & & & & & & 0.14 \\
      & & \ion{O}{4} & $1s^22s2p^2$($\frac{1}{2}$) &
$1s2s2p^3$($\frac{3}{2}$) & 1.0[-1] & 22.841 & & & & & & & 0.06 \\
$G$ & 22.871(4) & \ion{O}{5} & $1s^22s2p$(2) & $1s2s2p^2$(2) & 3.6[-6] & 22.882 &
22.877 & & & & & & 0.30 \\
      &        & \ion{O}{5} & $1s^22s2p$(2) & $1s2s2p^2$(3) & 4.2[-6] & 22.881 &
22.875 & & & & & & 0.05 \\
      &        & \ion{O}{5} & $1s^22s2p$(2) & $1s2s2p^2$(1) & 3.2[-7] & 22.882 &
22.874 & & & & & & 0.04 \\
$H$  & 23.017(10) & \ion{O}{6} & 
$1s^22p$($\frac{3}{2}$) & $1s2s^2$($\frac{1}{2}$) 
& 3.6[-3] &
  23.055 & 23.015 & 23.00 & 23.031 & 23.08 & & & 0.06 \\
        &        & \ion{O}{6} & 
$1s^22p$($\frac{1}{2}$)& $1s2s^2$($\frac{1}{2}$) & 3.6[-3] &
                      23.053 & 23.012 && 23.028 & & & & 0.03 \\
$I$  & 23.071(4) & \ion{O}{3} & $1s^22s^22p^2$(2) & $1s2s^22p^3$(1) & 8.8[-2] &
                           23.066 & & & & & & 23.170(10) & 0.13\\
      &        & \ion{O}{3} & $1s^22s^22p^2$(1) & $1s2s^22p^3$(1) & 8.0[-2] &
23.065 & & & & & & & 0.07\\
      &        & \ion{O}{3} & $1s^22s^22p^2$(0) & $1s2s^22p^3$(1) & 1.2[-1] &
23.065 & & & & & & & 0.03\\
      &        & \ion{O}{3} & $1s^22s^22p^2$(2) & $1s2s^22p^3$(2) & 2.0[-1] &
23.074 & & & & & & & 0.03\\
      &        & \ion{O}{3} & $1s^22s^22p^2$(2) & $1s2s^22p^3$(3) & 9.4[-2] &
23.086 & & & & & & & 0.02\\
\enddata
\tablenotetext{a}{Present measurement, numbers in parentheses
are uncertainties in m{\AA}.}
\tablenotetext{b}{Theoretical absorption oscillator strength of the transition
  with FAC, $a$[$b$] denotes $a\times 10^{b}$.}
\tablenotetext{c}{Configuration interaction calculation with FAC.}
\tablenotetext{d}{Configuration interaction with second-order MBPT
correction.}
\tablenotetext{e}{\citet{Gabriel72}}
\tablenotetext{f}{\citet{Vainshtein78, Safronova79}}
\tablenotetext{g}{\citet{Chen85, Chen86, Chen88}}
\tablenotetext{h}{\citet{Behar02}}
\tablenotetext{i}{Values for \ion{O}{5} and \ion{O}{6} are \textit{Chandra} 
  measurements adjusted for flow velocities from
  \citet{Kaastra04}, those for lower charge states are from Kaastra (private
  communications).}
\end{deluxetable}

\clearpage
\begin{deluxetable}{ *{6}{c} }
\scriptsize
\tablecaption{\label{tab:lines} Best estimate of wavelengths for the strongest
   transitions of \ion{O}{3} -- \ion{O}{6}.}
\tablehead{
\colhead{Ion} &
\colhead{Lower($J$)} &
\colhead{Upper($J$)} &
\colhead{$\lambda_{\mbox{exp}}$({\AA}) \tablenotemark{a}} &
\colhead{Label\tablenotemark{b}} &
\colhead{Index\tablenotemark{c}}
}
\startdata
\ion{O}{6} & $1s^22s$($\frac{1}{2}$) & $1s2s2p$($\frac{1}{2}$,$\frac{3}{2}$) &
22.374(8) & $u$,$v$ & $C$ \\
\ion{O}{6} & $1s^22p$($\frac{1}{2}$,$\frac{3}{2}$) & $1s2s^2$($\frac{1}{2}$) &
23.017(20) & $o$,$p$ & $H$ \\
\ion{O}{5} & $1s^22s^2$(0) & $1s2s^22p$(1) &
22.370(10) & $\beta$ & $C$ \\
\ion{O}{5} & $1s^22s2p$(0,1,2) & $1s2s2p^2$(0,1,2) &
22.449(8)  & & $D$ \\
\ion{O}{5} & $1s^22s2p$(2) & $1s2s2p^2$(2) &
22.871(5)  & & $G$ \\
\ion{O}{4} & $1s^22s^22p$($\frac{1}{2}$,$\frac{3}{2}$) &
$1s2s^22p^2$($\frac{1}{2}$,$\frac{3}{2}$) &
22.741(5) & & $E$\\
\ion{O}{4} & $1s^22s2p^2$($\frac{5}{2}$,$\frac{3}{2}$) & $1s2s2p^3$($\frac{3}{2}$)&
22.836(5) & & $F$\\
\ion{O}{3} & $1s^22s^22p^2$(1,2) & $1s2s^22p^3$(1) &
22.071(6) & & $I$\\
\enddata
\tablenotetext{a}{Present estimate, numbers in parentheses
are uncertainties in m{\AA}.}
\tablenotetext{b}{{L}ine labeling convention introduced by \citet{Gabriel72}.}
\tablenotetext{c}{The line feature index in Table~\ref{tab:id} to which the
   transitions belong.}
\end{deluxetable}

\clearpage

\clearpage

\end{document}